# $^{53}$Mn-$^{53}$Cr chronology and $\varepsilon^{54}$Cr-$\Delta^{17}$O genealogy of Erg Chech 002: the oldest andesite in the Solar System

Aryavart **Anand**[a*] (aryavart.anand@geo.unibe.ch)
Pascal M. **Kruttasch**[a] (pascal.kruttasch@geo.unibe.ch)
Klaus **Mezger**[a] (klaus.mezger@geo.unibe.ch)

[a] Institut für Geologie, Universität Bern, Baltzerstrasse 1+3, 3012 Bern, Switzerland

Corresponding author:
Aryavart Anand
(aryavart.anand@geo.unibe.ch)

**Abstract**

The meteorite sample Erg Chech (EC) 002 is the oldest felsic igneous rock from the Solar System analysed to date and provides a unique opportunity to study the formation of felsic crusts on differentiated protoplanets immediately after metal-silicate equilibration or core formation. The extinct $^{53}$Mn-$^{53}$Cr chronometer provides chronological constraints on the formation of EC 002 by applying the isochron approach using chromite, metal-silicate-sulphide and whole-rock fractions as well as ''leachates'' obtained by sequential digestion of a bulk sample. Assuming a chondritic evolution of its parent body, a $^{53}$Cr/$^{52}$Cr model age is also obtained from the chromite fraction. The $^{53}$Mn-$^{53}$Cr isochron age of 1.73 ± 0.96 Ma (anchored to D'Orbigny angirte) and the chromite model age constrained between $1.46^{+0.78}_{-0.68}$ and $2.18^{+1.32}_{-1.06}$ Ma after the formation of calcium-aluminium-rich inclusions (CAIs) agree with the $^{26}$Al-$^{26}$Mg ages (anchored to CAIs) reported in previous studies. This indicates rapid cooling of EC 002 that allowed near-contemporaneous closure of multiple isotope systems. Additionally, excess in the neutron-rich $^{54}$Cr (nucleosynthetic anomalies) combined with mass-independent isotope variations of $^{17}$O provide genealogical constraints on the accretion region of the EC 002 parent body. The $^{54}$Cr and $^{17}$O isotope compositions of EC 002 confirm its origin in the ''non-carbonaceous'' reservoir and overlap with the vestoid material NWA 12217 and anomalous eucrite EET 92023. This indicates a common feeding zone during accretion in the protoplanetary disk between the source of EC 002 and vestoids. The enigmatic origin of iron meteorites remains still unresolved as EC 002, which is more like a differentiated crust, has an isotope composition that does not match known irons meteorite groups that were once planetesimal cores.

## 1. Introduction

The majority of achondrites in our current meteorite collections are either basaltic or close to chondritic in composition. Achondrites with Si-rich evolved compositions (andesitic or trachyandesitic) are rare and include Graves Nunataks (GRA) 06128 and 016129 (Shearer *et al.*, 2010), ALM-A (Bischoff *et al.*, 2014), NWA 11119 (Srinivasan *et al.* 2018) and Erg Chech (EC) 002 (Barrat *et al.*, 2021). Among these, EC 002 represents the oldest felsic igneous rock from the Solar System analysed to date (Barrat *et al.*, 2021; Fang *et al.*, 2022). It is a 25% partial melting product of a chondritic source and provides a unique opportunity to study the formation of felsic crusts formed on a differentiated protoplanet immediately after metal-silicate equilibration or core formation (Barrat *et al.*, 2021). Magmatic iron meteorites that represent the cores of differentiated protoplanets have been extensively studied using $^{182}$Hf-$^{182}$W chronometry (e.g., Kruijer *et al.*, 2017) and more recently by $^{53}$Mn-$^{53}$Cr systematics (Anand *et al.*, 2021a). Both these chronometers constrain the timing of metal segregation in magmatic iron meteorite parent bodies to within 2.5 Ma after the formation of Ca-Al-rich inclusions (4567.18 ± 0.50 Ma, Amelin *et al.*, 2010). The core formation ages in magmatic iron meteorite parent bodies agree with the $^{26}$Al-$^{26}$Mg isochron for EC 002 which yields an age of 1.80 ± 0.01 Ma after CAIs, assuming a homogeneous distribution of $^{26}$Al in the early Solar System with an initial $^{26}$Al/$^{27}$Al ratio of 5.23×10$^{-5}$ (Fang *et al.*, 2022; Jacobsen *et al.*, 2008). The short-lived $^{53}$Mn-$^{53}$Cr ($t_{1/2}$ = 3.7 ± 0.4 Ma; Honda *et al.*, 1971) chronometer is another powerful tool to constrain the time and duration of early Solar System processes including accretion, differentiation, metamorphism and subsequent cooling (e.g., Shukolyukov and Lugmair, 2006; Trinquier *et al.*, 2008b; Zhu *et al.*, 2021; Anand *et al.*, 2021a, b) and can help to refine the timing and extent of the formation process of a primitive igneous crust on the EC 002 parent body.

Manganese and Cr are minor elements in EC 002 mineral phases, and these phases have variable and relatively high Mn/Cr ratios, making them suitable for dating by the isochron method. Additionally, the $^{53}$Cr/$^{52}$Cr ratio can be used to obtain ''model ages'' for chromite which has a Mn/Cr ratio near zero (Anand *et al.*, 2021a, b). The combination of $^{54}$Cr/$^{52}$Cr (expressed as ε$^{54}$Cr) and O isotopes (expressed as Δ$^{17}$O), which were heterogeneously distributed in the early Solar System (Warren, 2011), can be used to identify the isotopic source reservoir of the parent body and potentially relate EC 002 to other achondritic meteorites. Previous studies have established genealogical connections between different meteorite groups in ε$^{54}$Cr vs. Δ$^{17}$O space, such as between IIIAB irons, main group pallasites and HEDs (e.g.,

Wasson, 2013), IVA irons and L/LL chondrites (e.g., Anand *et al.* 2021c) and IIG and IIAB irons (Anand *et al.*, 2022). Thus $^{53}$Mn-$^{53}$Cr systematics constrain the timing of felsic crust formation, and the $\varepsilon^{54}$Cr-$\Delta^{17}$O data provide genealogical constraints on the EC 002 parent body and its source reservoir.

## 2. Analytical Methods

Approximately ~2 g EC 002 meteorite fragments were obtained from Decker Meteorite Shop, Germany (www.meteorite-museum.de). After cleaning in an ultrasonic bath with ethanol, the fragments were crushed into a fine powder using an agate mortar and pestle. This whole-rock powder was successively used to perform three experiments on EC 002 as shown in Fig. 1. In the first experiment, ~20 mg of whole-rock powder was digested in a Parr® bomb using a concentrated HF-HNO$_3$ mixture. This fraction is labelled as ''whole-rock''. The second experiment involved ~100 mg whole-rock powder subjected to a concentrated HF-HNO$_3$ mixture on a hot plate at 130 °C for 24 hours and split into two fractions: (1) completely digested sulphide-metal-silicate fraction, labelled as ''silicates'' and (2) refractory phase, subsequently digested in a Parr® bomb and labelled as ''chromite''. The third experiment involved a sequential digestion procedure (adopted from Yamakawa *et al.*, 2010) using ~1.5 g whole rock material to obtain data from phases with various Mn/Cr ratios (Rotaru *et al.*, 1992). The whole procedure involved 6 leaching steps labelled as ''leachate 1'' to ''leachate 6'' and summarized in Table 1. Before chemical separation, an aliquot from digested fractions of all three experiments was diluted in 10 ml 0.5 M HNO$_3$ to produce a 10 ppb Cr solution. These aliquots were used to determine Cr, Mn, and Fe concentrations using a 7700x Agilent ICP-MS at the Institute of Geography, University of Bern. Uncertainties on $^{55}$Mn/$^{52}$Cr and Fe/Cr ratios are reported in 2 standard errors of the replicate measurements (n = 5) and remained <5 % for all the samples.

The procedure for Cr purification was adopted from Schoenberg *et al.* (2016). It includes three steps of a combination of cation-anion exchange chromatography modified after Schoenberg and von Blanckenburg (2005) (column 1), Trinquier *et al.* (2008a) and Yamakawa *et al.* (2009) (columns 2 and 3). In brief, an aliquot containing ~15 µg of Cr from each sample was taken up in 1 ml 6 M HCl and loaded onto the first 7.5 mL Spectrum® polypropylene column containing 2 ml anion resin (BioRad® AG 1X8 100-200 mesh size). The Cr eluate from the first column was dried down, redissolved in 400 µl 6 M HCl, equilibrated on a hotplate set to 130 °C for ~30 min one day before the chemical separation and stored at room temperature overnight. The next day, the sample was re-equilibrated on a hot plate at 130 °C for one hour, diluted with 2 ml of MilliQ® water to obtain 2.4 ml 1 M HCl and loaded onto the second column filled with 2 ml cation resin (BioRad® AG 50W-X8 200-400 mesh size). The second column separation step produced a solution with mostly Cr, but incompletely separated from Ti and V. The eluate from the second column was dried down on a hot plate (at 90 °C), taken up in 0.5 ml conc. $HNO_3$ and dried down again immediately to transform the samples into nitrate form. The residue was redissolved in 3 ml 0.4 M $HNO_3$ for 30 min on a hotplate at 80 °C and let react cold for 5 days for the production of chloro-aquo complexes. Afterwards, each sample was loaded onto the third column containing 0.5 ml BioRad® AG 50W-X8 200-400 mesh resin in order to obtain a clean Cr separate, free of Ti and V. The matrix was eluted in 8 ml 0.5 M HF and 9.5 ml 1 M HCl, and Cr was collected in 8 ml 4 M HCl. Finally, the Cr separate was redissolved in 100 µl conc. $HNO_3$ and dried immediately at 130 °C on a hotplate. This step was repeated 2 times to remove the residual organics from the column chemistry. Typical recovery of Cr was in excess of 80% for the whole column chemistry. Total chemistry blanks were below 20 ng, which are negligible compared to the µg range of Cr processed through the columns.

After column chemistry, purified Cr samples were analysed on a Thermal Ionization Mass Spectrometer (Thermo Scientific TRITON Plus) at the Institute of Geological Sciences, University of Bern. Two to three µg of Cr from each sample was loaded on multiple filaments and measured at a $^{52}Cr$ signal intensity between 7 and 10 V ($10^{-11}$ Ω resistor). Intensities of $^{50}Cr$, $^{51}V$, $^{52}Cr$, $^{53}Cr$, $^{54}Cr$, $^{55}Mn$ and $^{56}Fe$ were measured on the Faraday cups L3, L2, L1, C, H1, H2 and H3, respectively (Trinquier *et al.,* 2008a). Alignment of all the Cr peaks (peak scan) was ensured before each measurement and the peak center was monitored on $^{53}Cr$ in the center cup. Isobaric interference of $^{54}Fe$ on $^{54}Cr$ was corrected by monitoring $^{56}Fe$. The isotope $^{51}V$ was measured and $^{49}Ti$ was monitored to correct for isobaric interferences on $^{50}Cr$. However, the $^{49}Ti$ and $^{51}V$ intensities remained indistinguishable from background intensities for all samples, verifying the successful separation of V and Ti from Cr during column chromatography. A typical run for a single filament load consisted of 24 blocks with 20 cycles each (integration time = 8.389 s), obtained in static acquisition mode. Gain calibration was done once, at the beginning of every analytical session. The baseline was measured, and the amplifiers were rotated after every block (baseline = 30 cycles, each of 1.05 s). The Cr standard reference material NIST SRM 979, was used as a terrestrial reference material. The $^{53}Cr/^{52}Cr$ and $^{54}Cr/^{52}Cr$ ratios were normalized to $^{52}Cr/^{50}Cr$ = 19.28323 (Shields *et al.,* 1966) by applying the exponential mass fractionation law and are reported as $\varepsilon^i$Cr, where $\varepsilon^i$Cr = ([$^i$Cr/$^{52}$Cr]$_{sample}$ / [$^i$Cr/$^{52}$Cr$_{NIST\ SRM\ 979}$] - 1) x $10^4$ (i = 53 or 54).

The $\varepsilon^i$Cr reported for any one sample represents the mean of the replicate measurements (n = 3-5, Tables 1 and Appendix). The replicate measurements for each sample are used to determine the external precision reported as 2 standard errors. The isotope compositions of each sample are reported relative to the mean value of the standard reference material (NIST SRM 979) measured along with the samples in each measurement session (single turret). The external precision (2 standard errors, n = 8) for the standard reference material (NIST SRM 979) in each measurement session was ~0.05 for $\varepsilon^{53}$Cr and ~0.11 for $\varepsilon^{54}$Cr.

The model age for EC 002 is determined by comparing the $^{53}$Cr/$^{52}$Cr isotopic ratio of the chromite fraction to the theoretical $^{53}$Cr/$^{52}$Cr evolution of the chondritic reservoir (Anand *et al.*, 2021a) and assuming a homogeneous abundance of $^{53}$Mn in the early Solar System (e.g., Trinquier *et al.*, 2008b), known abundances of $^{53}$Mn and $^{53}$Cr at the beginning of the Solar System (or any point in time thereafter), an estimate for the $^{55}$Mn/$^{52}$Cr in the relevant reservoir and known decay constant of $^{53}$Mn. The spallogenic Cr contributions in the $\varepsilon^{53}$Cr and $\varepsilon^{54}$Cr isotopic compositions are corrected using $^{53}$Cr and $^{54}$Cr production rates in Fe targets from the Grant iron meteorite (2.9 × $10^{11}$ atoms/Ma, Birck and Allegre, 1985) and the relationship described in Trinquier *et al.* (2007), assuming a spallation depth profile similar to that in Graf and Marti (1995). The cosmic-ray exposure age for EC 002 ($^3$He exposure age of 26.0 ± 1.6 Ma) is taken from Barrat *et al.* (2021). Isochron regressions (Fig. 3a) are generated using IsoplotR (Vermeesch, 2018) with a maximum likelihood regression model. The O isotope data for EC 002 genealogy are taken from Gattacceca *et al.* (2021).

### 3. Results

The Cr isotopic composition and $^{55}$Mn/$^{52}$Cr and Fe/Cr ratios of the EC 002 fractions and reference materials are provided in Table 2. The Cr isotopic compositions of the terrestrial reference material OKUM are indistinguishable from typical terrestrial values (Zhu *et al.,* 2021 and references therein) and Cr isotopic compositions of Allende are consistent with literature data (Trinquier *et al.*, 2008b) confirming the analytical accuracy of the Cr isotope data. The $^{55}$Mn/$^{52}$Cr ratio (= 1.04 ± 0.05) determined for the whole rock EC 002 agrees with $^{55}$Mn/$^{52}$Cr = 1.09 (MnO = 0.46 wt. %, Cr$_2$O$_3$ = 0.57 wt. %), determined in Carpenter *et al.* (2021), but is lower than that determined in Barrat *et al.* (2021), who obtained $^{55}$Mn/$^{52}$Cr = 1.51 (MnO = 0.47 wt. %, Cr$_2$O$_3$ = 0.42 wt. %). The difference in $^{55}$Mn/$^{52}$Cr ratio relative to Barrat *et al.* (2021)

might be related to mineral scale sample heterogeneity. Due to the larger quantity of EC 002 analyzed in the present study, the $^{55}$Mn/$^{52}$Cr ratio (= 1.04 ± 0.05) reported here is the most representative to date. The proportion of Mn and Cr extracted in each sequential digestion step is given in Table 1. The Cr isotopic compositions of leachate 1 could not be measured due to the low amount of Cr leached in this step (< 0.03 %, Table 1). The Cr isotopic composition of all the fractions are corrected for spallogenic Cr contributions and the measured and corrected ratios are listed in Table 2. The absolute variation in Cr isotope composition in all the EC 002 fractions is 1.07ε in $^{53}$Cr and 0.58ε in $^{54}$Cr, which reduces to 1.04ε in $^{53}$Cr and 0.46ε in $^{54}$Cr after correction for spallogenic Cr contribution. The correction for spallogenic Cr contribution appears to be significant in leachates 2, 3 and 4 which is evident in the correlation in the ε$^{54}$Cr vs. Fe/Cr plot (Fig. 2a). However, Cr isotopic compositions of leachate 2 remains to be distinct from the Cr isotopic compositions of the other leachates in the ε$^{53}$Cr vs. ε$^{54}$Cr plot (Fig. 2b) after correction for spallogenic Cr. In order to obtain an age for EC 002, all the analyzed fractions, whole rock, silicates, chromite and leachates L2-L7, are plotted in an isochron diagram in Fig. 3a. The slope of this $^{53}$Mn-$^{53}$Cr isochron ($^{53}$Mn/$^{55}$Mn = 4.76 ± 0.39 x 10$^{-6}$) when anchored to the D'Orbigny angrite (Pb-Pb age of 4563.37 ± 0.25 Ma (Brennecka and Wadhwa, 2012) and initial $^{53}$Mn/$^{55}$Mn ratio of 3.24 ± 0.04 x 10$^{-6}$ (Glavin *et al.*, 2004)), yields an age of 1.73 ± 0.96 Ma after CAI formation (Fig. 3a). The first dissolution steps release the labile Mn and Cr components from the sample. This is also the material that is easiest modified by alteration on the parent body and weathering in the terrestrial environment. Thus Yamakawa *et al.* (2010) eliminated the data points for leachates L1-L3 from the isochron produced for ureilite NWA 766 on the suspicion of being affected by parent body or terrestrial alteration. This choice is supported by the observation that these points do not fall on the isochron defined by the more resistant components. In contrast the data points for leachates L2 and L3 of EC 002 fall on the isochron within analytical uncertainties, which attests to the pristine character of this sample and the absence of post-formation alteration of the Mn-Cr systematics. Therefore, these early leaching steps are included in the regression of the isochron.

In addition to the isochron age, a model age of $2.18^{+1.32}_{-1.06}$ Ma is derived for the chromite fraction relative to the CAI formation age, assuming a chondritic $^{55}$Mn/$^{52}$Cr = 0.74 of the source reservoir (Zhu *et al.*, 2021), a Solar System initial ε$^{53}$Cr = -0.30 (Anand *et al.*, 2021a) and a canonical $^{53}$Mn/$^{55}$Mn = 6.28 x 10$^{-6}$ (Trinquier *et al.*, 2008b) (Fig. 2b).

## 4. Discussion

**$^{53}$Mn-$^{53}$Cr chronology of Erg Chech 002**

The $^{53}$Mn-$^{53}$Cr isochron age of 1.73 ± 0.96 Ma obtained for EC 002 using its chromite, silicates and whole-rock fractions and sequential digestion leachates is calculated relative to the angrite D'Orbigny as an anchor. With another commonly used angrite anchor, LEW 86010 (Pb-Pb age of 4557.5 ± 0.3 Ma (Kleine *et al.*, 2012) and initial $^{53}$Mn/$^{55}$Mn ratio of 1.25 ± 0.07 x 10$^{-6}$ (Lugmair and Shukolyukov, 1998)), the age would be 2.46 ± 0.96 Ma after CAIs i.e., 0.73 Ma younger. However, both ages are analytically unresolvable given the ~1 Ma uncertainty associated with the age estimates. A more significant difference in isochron ages based on the choice of anchor is reported in the $^{26}$Al-$^{26}$Mg chronometry of EC 002. Fang *et al.* (2022) reported an initial $^{26}$Al/$^{27}$Al ratio of (8.89 ± 0.09) × 10$^{-6}$ for an isochron from EC 002 obtained using mineral separates (pyroxenes and plagioclases) and analysed with multi-collector inductively coupled plasma mass spectrometer (MC-ICP-MS). The initial $^{26}$Al/$^{27}$Al ratio of (8.89 ± 0.09) × 10$^{-6}$ yields an age of 0.60 ± 0.01 Ma, when D'Orbigny angrite is used as an anchor (Brennecka and Wadhwa, 2012; Schiller *et al.*, 2015) and 1.80 ± 0.01 Ma when anchored to the first-formed refractory solids (CAIs) in the Solar protoplanetary disk that are characterized by a canonical $^{26}$Al/$^{27}$Al ratio of 5.23×10$^{-5}$ (Jacobsen *et al.*, 2008). In an earlier study, Barrat *et al.* (2021) analyzed the plagioclase and pyroxene crystals in EC 002 using an ion microprobe and reported an initial $^{26}$Al/$^{27}$Al ratio of (5.72 ± 0.07) × 10$^{-6}$, which translates into an age of 1.16 ± 0.01 Ma when anchored to the D'Orbigny angrite and 2.26 ± 0.01 Ma when anchored to the canonical $^{26}$Al/$^{27}$Al ratio derived from CAIs. First, there is a ~0.5 Ma discrepancy between the EC 002 initial $^{26}$Al/$^{27}$Al ratios reported in Barrat *et al.* (2021) and Fang *et al.* (2022). This has been explained by Fang *et al.* (2022) as a result of the much faster diffusion of Mg in plagioclase than in pyroxene (Van Orman *et al.,* 2014; Zhang *et al.,* 2010) because of which the bulk plagioclase-rich fractions analysed by MC-ICP-MS better represent a closed system for Mg diffusion than the core of the plagioclase crystals analysed using the ion microprobe. Hence, the MC-ICP-MS $^{26}$Al-$^{26}$Mg age is closer to the age of crystallization. Second, there is a discrepancy of ~1 Ma in $^{26}$Al-$^{26}$Mg chronometry of EC 002 between the choice of anchor (CAIs and D'Orbigny angrite). Previous studies have explained the discrepancy between $^{26}$Al-$^{26}$Mg ages when anchored to angrites and CAIs by suggesting a lower initial $^{26}$Al/$^{27}$Al of the angrite parent body precursor material at the time of CAI formation (e.g., Larsen *et al.,* 2011; Schiller *et al.,* 2015). However, this possible heterogeneity has been questioned by other studies (Wasserburg *et al.,* 2012; Kita *et al.,* 2013; see also Fukuda *et al.,* 2022 for detailed discussion). The robustness of $^{26}$Al-$^{26}$Mg system in D'Orbigny has also been questioned because of the complex petrologic history of this angrite based on its apparent disturbed Sm-Nd systematics (Sanborn et al., 2015; Sanborn et al., 2019). In view of these arguments, the distribution of $^{26}$Al in the early Solar System (i.e., homogeneous vs heterogeneous) is still an open question. In the present study, the $^{26}$Al-$^{26}$Mg crystallization age of 1.80 ± 0.01 Ma (Fang et al., 2021) obtained using CAI initial $^{26}$Al/$^{27}$Al = 5.23×10$^{-5}$ is used in the rest of the discussion as also done in both the $^{26}$Al-$^{26}$Mg studies conducted on EC 002 (Barrat et al., 2021; Fang et al., 2021).

Although the use of anchors in short-lived isotope systematics introduces systematic uncertainties due to age shifts, nevertheless, both $^{53}$Mn-$^{53}$Cr and $^{26}$Al-$^{26}$Mg (when anchored to CAIs) isochron ages are concordant within analytical uncertainties. The $^{53}$Mn-$^{53}$Cr and $^{26}$Al-$^{26}$Mg (when anchored to CAIs) isochron ages are also in agreement with the model age of $2.18^{+1.32}_{-1.06}$ Ma after CAIs, determined using EC 002 chromite fraction (Fig. 6), which requires estimates for the Solar System initial $^{53}$Mn and $^{53}$Cr abundances and $^{55}$Mn/$^{52}$Cr of the source reservoir. The Solar System initial $\varepsilon^{53}$Cr = -0.30 used in the present study was determined in Anand et al. (2021a) by calibrating the $^{53}$Mn-$^{53}$Cr core formation ages with the previously determined $^{182}$Hf-$^{182}$W ages (Kruijer et al., 2017) of magmatic iron meteorite groups. Moreover, $^{55}$Mn/$^{52}$Cr of the EC 002 source reservoir from which the chromite formed is assumed to be chondritic. Although a chondritic $^{55}$Mn/$^{52}$Cr is a good estimate for the EC 002 parent body (Barrat et al., 2021), the source reservoir for the EC 002 meteorite is a fractionated product after metal-silicate separation on the parent body and hence, may have an evolved $^{55}$Mn/$^{52}$Cr composition. Thus, the chromite model age of $2.18^{+1.32}_{-1.06}$ Ma after CAIs, determined assuming a chondritic $^{55}$Mn/$^{52}$Cr ratio for the source of the chromite in EC 002 provides an upper limit (younger limit). To constrain the lower limit (older limit), the $^{55}$Mn/$^{52}$Cr composition of the EC 002 parent body can be assumed to be the whole rock $^{55}$Mn/$^{52}$Cr (= 1.04 ± 0.05) composition of EC 002. The chromite model age determined in this case would be $1.46^{+0.78}_{-0.68}$ Ma after CAIs, which appears to be 0.73 Ma older (Fig. 3b), however, still in agreement within uncertainties, given current analytical resolution. In summary, consideration of the full range of the chromite model ages indicates the crystallization of EC 002 between $1.46^{+0.78}_{-0.68}$ Ma and $2.18^{+1.32}_{-1.06}$ Ma after CAIs.

Figure 6 presents a compilation of $^{53}$Mn-$^{53}$Cr and $^{26}$Al-$^{26}$Mg chronometry of EC 002. For both the $^{53}$Mn-$^{53}$Cr and $^{26}$Al-$^{26}$Mg (when anchored to the CAIs) systems, the ages are concordant with the $^{53}$Mn-$^{53}$Cr chromite model ages. This provides confidence over the estimates of $^{53}$Mn and $^{53}$Cr abundances and $^{55}$Mn/$^{52}$Cr of the source reservoir used for chromite model age determination and indicates rapid cooling of EC 002 on its parent body at around the time of isotopic closures of both $^{53}$Mn-$^{53}$Cr and $^{26}$Al-$^{26}$Mg systems. The internal isochrons date the crystallization of the parent melt of EC 002 and the two isotopic systems did not re-equilibrate after initial and near-simultaneous closure. This result confirms the thermal history of EC 002 inferred by Barrat *et al.* (2021) using experimental data. The authors modelled the zoning profile of Mg# across a small xenocryst and estimated the cooling rate to be about 5°C/a between 1,200 and 1,000°C. Furthermore, only cristobalite and tridymite (PO) are observed in EC 002 with a total absence of quartz. Cristobalite transforms to tridymite above 900 °C, but it also transforms to quartz easily. The lack of this transformation to quartz implies rapid cooling at a rate of >0.1 to 1°C/d below 900°C. The initial slow cooling allowed the EC 002 to acquire its medium-grained groundmass texture and later rapid cooling enabled multiple isotopic systems to close within the given resolution. The change in the cooling rate has been explained by an impact that ejected the EC 002 rock from its parent body (Barrat *et al.*, 2021). The rapid cooling around the time of isotopic closures of multiple isotopic systems, no subsequent partial re-equilibration event and insignificant secondary processes that may have disturbed the isotopic systems make EC 002 itself a potential relative time anchor (Sanborn *et al.*, 2019).

Recently, Zhu *et al.* (2022) have reported $^{53}$Mn-$^{53}$Cr systematics for bulk and mineral components from Erg Chech 002 and observed mineral scale sample heterogeneities in EC 002, suggesting a heterogeneous mantle sources of $\varepsilon^{54}$Cr in the EC 002 parent body. They determined a $^{53}$Mn-$^{53}$Cr isochron age of 0.7 ± 0.6 Ma after CAIs for EC 002 crystallization, considering only the cogenetic matrix fractions with 'overlapping' $\varepsilon^{54}$Cr values. This age is ~1 Ma older than the $^{53}$Mn-$^{53}$Cr ages (both isochron and chromite model ages) reported in the present study as well as the $^{26}$Al-$^{26}$Mg ages reported in the previous studies (Barrat *et al.,* 2021; Fang *et al.,* 2021). The inconsistent age reported in Zhu *et al.* (2022) is most likely due to a large fraction of xenolithic material with anomalous $\varepsilon^{54}$Cr incorporated in their studied sample mass. The variation in the $\varepsilon^{54}$Cr data for individual components (whole rock, chromite, silicates and sequential digestion leachates) reported in Table 2 is much lower than the $\varepsilon^{54}$Cr data for mineral components from Erg Chech 002 reported in Zhu *et al.* (2022) although the total samples mass processed in the present study is ~10 times larger than that in Zhu *et al.* (2022).

**$\varepsilon^{54}$Cr-$\Delta^{17}$O genealogy of Erg Chech 002**

Nucleosynthetic isotope variations of $^{54}$Cr and $^{17}$O are powerful tools that can provide insights into the accretion region of EC 002 parent body (e.g., Clayton and Mayeda, 1996; Greenwood et al., 2012; Shukolyukov and Lugmair, 2006; Trinquier *et al.*, 2007). In addition, these isotopes can be used to deduce an affinity of EC 002 with other achondrites and chondrites. The $\varepsilon^{54}$Cr values obtained for EC 002 whole rock and its sequential digestion leachates after correction for spallogenic Cr contribution are within the analytical uncertainties of each other, except for the leachate 4 (which hosts only 0.27% Cr of the WR). This indicates homogeneity of non-radiogenic Cr isotopes in the mineral phases on a larger scale (Table 2, Fig. 2b). An isotopic equilibrium of $\varepsilon^{54}$Cr in mineral phases is expected in EC 002 if they all formed from the same melt since the effects of melting and planetary differentiation generally leads to homogenization of the isotopic signatures within a planetary body (can also be a disequilibrium process, see van Kooten *et al.*, 2017). However, a distinct $\varepsilon^{54}$Cr isotopic composition of leachate 4 that represents silicate fraction of the EC 002 (Rotaru *et al.*, 1992) most likely reflects contribution from the olivine and pyroxene xenocrysts present in EC 002 (Barrat *et al.*, 2021). Nevertheless, this contribution from xenocrysts does not affect the isochron and by extension the chronology of the EC 002 (Fig. 3a); and hence appears to be insignificant within uncertainties in the Cr isotopic budget. Alternatively, the high $^{54}$Cr corrected for spallogenic $^{54}$Cr derived from Fe could reflect the fractionation of Fe from Cr during leaching.

The average $\varepsilon^{54}$Cr of the EC 002 fractions (= -0.65 ± 0.10) confirms its affinity to the 'non-carbonaceous (NC)' reservoir (Trinquier *et al.*, 2007, 2009) as speculated in Barrat *et al.* (2021), who determined a non-carbonaceous chondrite-like negative Tm anomaly (Tm/Tm* = 0.973, where Tm* is the expected Tm concentration for a smooth CI- normalized REE pattern). The $\mu^{148}$Nd value of EC 002 reported in Fang *et al.* (2022) is also distinctly lower than any CC meteorite and falls within the NC region. The O isotope compositions of EC 002 plot distinctly away from other andesitic and trachyandesitic achondrites, GRA 06128, GRA 016129, ALM-A and NWA 11119, and lie between the fields for the majority of eucrites and angrites (Gattacceca *et al.*, 2021; Fig. 4). The O isotopic compositions are also in close vicinity to anomalous eucrites, Bunburra Rockhole, Emmaville, Asuka 881394 and EET 92023 and ungrouped achondrites, NWA 12217, NWA 12562 and NWA 12319 (Fig. 4). Interestingly, one of the O isotope data points of EC 002 is indistinguishable from NWA 12217. The coupled $\Delta^{17}$O and $\varepsilon^{54}$Cr systematics of EC 002 could bring more clarity to this genealogical observation made solely based on O isotope compositions. In $\varepsilon^{54}$Cr-$\Delta^{17}$O space, EC 002 appears to be distinct from the field of angrites, and normal and anomalous HEDs but a genetic link with meteorites NWA 12217 and EET 92023 is possible. The $\varepsilon^{54}$Cr and $\Delta^{17}$O isotope composition of EC 002 overlaps with both NWA 12217 and EET 92023 (Fig. 5). As reported in Gattacceca *et al.* (2021), the unbrecciated anomalous eucrite EET 92023 is mineralogically quite different from EC002 in several aspects, most notably the prevalence of highly sodic plagioclase and lower FeO/MnO ratios in pyroxenes. The ungrouped achondrites NWA 12217, NWA 12562 and, NWA 12319 are reported to be sourced from a fully differentiated parent body that is likely to be a V-type asteroid or vestoid (Vaci *et al.*, 2021). The heavier $\Delta^{17}$O composition of NWA 12217 compared to NWA 12562 and NWA 12319 and main group normal HEDs is explained as having been influenced by mixing with an ordinary chondrite component (Vaci *et al.*, 2021), possibly by an impact event. If EC 002 is sourced from the same vestoid reservoir as these achondrites, then the $\Delta^{17}$O composition of EC 002, which is similar to that of NWA 12217, would also require addition of a higher $\Delta^{17}$O and $\varepsilon^{54}$Cr component to form EC 002 (Fig. 5). In the case of EET 92023, admixing of an even higher $^{54}$Cr component would be required if the sample originally formed on Vesta. However, Yamaguchi *et al.,* (2017) argued that a significant amount of kamacite, taenite and platinum group elements (PGEs) in EET 92023, which are uncommon in unbrecciated crystalline eucrites, indicates incorporation of material most likely from an iron meteorite rather than a chondrite material. The dunite and lherzolite cumulates, NWA 12217, NWA 12562 and NWA 12319, characterize an olivine-rich mantle from a vestoid body that experienced core formation and silicate differentiation and may explain the low HSE abundances and oxygen fugacity required for the composition of EC 002 (Nicklas *et al.,* 2021). Alternatively, if EC 002 did not originate from the same vestoid parent body, it could have formed on a different parent body that accreted from material sourced from an overlapping feeding zone in the protoplanetary disk. Chronologically, the first eucrite melts formed approximately 3 to 5 Ma after CAIs (i.e., ∼1-3 Ma after EC 002 crystallisation) and peaked even later at about 11 Ma after CAI formation (Zhou *et al.,* 2013; Roszjar *et al.,* 2016). Thus, if EC 002 formed on a vestoid, the parent body must have accreted and differentiated in the first few million years of the Solar System. Early accretion and differentiation of the EC 002 parent body are consistent with a modelled accretion age for Vesta of 1.50 to 1.75 Ma after CAI (Mitchell *et al.,* 2021). The early ejection of EC 002, as observed in its cooling history, may have prevented it from destruction by the basaltic volcanism that occurred later on its parent body.

The crustal material, EC 002 and EET 92023, and mantle-derived samples, NWA 12217, NWA 12562 and NWA 12319, are difficult to genetically link to the existing collection of magmatic iron meteorite groups that represent iron cores of differentiated planetesimals. This is because of the rare occurrences of mineral phases such as oxides and silicates in iron meteorites that could be used for O and Cr isotope analyses. The IIIAB iron meteorite group has an O-isotopic composition close to that of the HEDs (Clayton and Mayeda, 1996). Figure 5 shows that in $\epsilon^{54}$Cr-$\Delta^{17}$O space, chromite fractions from IIIAB iron, Cape York lie in the HED field but not in the vicinity of EC 002 (Franchi *et al.* 2013; Anand et al. 2022). Magmatic iron meteorite groups such as IVA, IIAB and IIG also plot distinctly away from EC 002, NWA 12217 as well as normal HEDs (Anand *et al.* 2021c, 2022). Apart from the magmatic iron meteorites, Dey *et al.* (2022) have reported overlapping $\epsilon^{54}$Cr-$\Delta^{17}$O compositions of NWA 12217 and the main group pallasites Imilac, Hambleton and Brenham and hence these sample may also share the same heritage as EC 002 (Fig. 5). However, isotopic differences between metal and silicates in pallasites are also reported that imply that the silicate and metal portions of these meteorites may have distinct isotopic reservoirs (Windmill *et al.* 2022), possibly due to late accretion after core formation.

## 5. Conclusions and Outlook

Figure 6 provides a compilation of chronological data obtained for EC 002 (present study and Barrat *et al.* 2021) and the three largest magmatic iron meteorite groups, IIAB, IIIAB and IVA, within the NC reservoir (Kruijer *et al.* 2017; Anand *et al.*, 2021a). EC 002 represents the oldest magmatic rock dated so far, and it is also the only silicate material with a formation age contemporaneous with the core-mantle differentiation in the oldest planetesimals of the Solar System. EC 002 originated from the primitive igneous crust of an early accreted and differentiated planetesimal, the mantle of which may be represented by the ultramafic achondrites NWA 12217, NWA 12562 and NWA 12319 and the core represents some undiscovered magmatic iron meteorite group. Thus, the enigmatic origin of iron meteorites remains unresolved as EC 002, which is more like differentiated crust does not match known irons meteorite groups that were once cores and the silicate part belonging to these cores is still to be discovered.

The timing and duration of chondrule formation can provide key constraints and context to the spatio-temporal evolution of the earliest formed planetary bodies such as EC 002. The two commonly used chronometers for chondrule formation ages are the long-lived, absolute $^{207}$Pb-$^{206}$Pb system (e.g., Connelly et al., 2012; Bollard et al., 2017) and the short-lived $^{26}$Al-$^{26}$Mg system (e.g., Pape *et al.* 2019; Siron *et al.*, 2021, 2022). The $^{207}$Pb-$^{206}$Pb system indicates a protracted chondrule formation for ordinary chondrites spanning from 0 to 4 Ma after CAIs

(Connelly *et al.*, 2012; Bollard *et al.*, 2017), while the $^{26}$Al-$^{26}$Mg chondrule ages in unequilibrated ordinary chondrites indicate a restricted formation duration from 1.8 to 2.2 Ma after CAIs (Siron et al., 2021, 2022; Fukuda et al. 2022). The discrepancy between $^{207}$Pb-$^{206}$Pb and $^{26}$Al-$^{26}$Mg chronometers is to some extent due to the lack of a better choice of age anchor between absolute and relative ages (see Siron *et al.*, 2021 for more discussion). The difference may also occur because the U-Pb system dates the formation of chondrule precursor material and the $^{26}$Al-$^{26}$Mg system dates melt formation in the chondrules (e.g., Pape *et al.*, 2019).

If the $^{26}$Al-$^{26}$Mg chondrule ages date melt formation in the chondrules, then the EC 002 crystallization age and the core formation ages in the magmatic iron meteorite parent bodies either predate or are at the earlier side of the chondrule formation interval defined by the $^{26}$Al-$^{26}$Mg studies of chondrules in unequilibrated ordinary chondrites (e.g., Pape *et al.* 2019; Siron *et al.*, 2021, 2022). For comparison, the second oldest felsic igneous rock, NWA 11119 (Srinivasan *et al.* 2018), previously constrained the silica-rich volcanism in the early Solar System within the first 2.5-3.5 million years of the Solar System history which is after the peak of the chondrule formation interval (e.g., Pape *et al.* 2019; Siron *et al.*, 2021, 2022). The granitoids and andesitic fragments found within two chondrite regolith breccias Adzhi-Bogdo and Study Butte also yielded formation ages of at least 3.8 Ma after CAIs and hence, after the chondrule formation interval as well as chondrite parent body accretion (Sokol *et al.*, 2017). The EC 002 crystallization age and core formation ages in magmatic iron meteorite parent bodies predating the chondrule formation interval defined by the $^{26}$Al-$^{26}$Mg studies of chondrules in unequilibrated ordinary chondrites are in line with the argument that chondrule formation may not necessarily be an intermediate step on the way from dust to planets and early formed planetesimals may have played an active role in chondrule formation (e.g., Anand *et al.*, 2021a, b).

Chondrule formation models that link early formed planets to chondrule formation processes include collisions of planets or bow shocks caused by migrating planetesimals or early formed planets (Connolly and Jones, 2016 and references therein). Chondrule formation due to planetesimal bow shock mechanism involves early formed planetesimals (Mann *et al.*, 2016), and large planetesimals, with a diameter of at least ~1000 km. Other mechanisms that can explain chondrule formation by planetesimal collision could be impact splashes during planetesimal recycling (Lichtenberg *et al.*, 2018) and/or chondrule formation as a result of inefficient pairwise accretion, when molten or partly molten planetesimals ~30-100 km diameter, similar in size, collided at velocities similar to their two-body escape velocity ~100 m/s (Asphaug *et al.*, 2011). Since, EC 002 sample originates from a highly differentiated material, if such material was involved in chondrule formation by collisions, then it would have been hard to mix the differentiated material from metal core to silicate mantle to differentiated crust in such a way as to obtain chondritic abundances of highly siderophile and also highly incompatible elements in chondrites. Nevertheless, an experimental study by Faure (2020) showed that early silica crust formation (before complete differentiation) in planetesimals by metastable silica-rich liquid immiscibility or cristobalite crystallization can explain the possible origin of silica-rich chondrules. In a recent study, Sturtz *et al.* (2022) proposed a model of planetesimals formation and evolution for the parent bodies like that of EC 002. The authors showed that a protracted and continuous accretion of primitive chondritic material over a magma ocean generated due to early accretion of the parent body can lead to the preservation of a few km thick chondritic crust. During accretion of this material, the heat released by the radioactive decay of extant $^{26}$Al can induce partial melting beneath the chondritic crust leading to the production of rocks such as EC 002 within the first 1-2 Myr of the Solar System.

**Declaration of Competing Interest**

The authors declare that they have no known competing financial interests or personal relationships that could have appeared to influence the work reported in this paper.

**Acknowledgements**

This study was partially funded by the 'Swiss Government Excellence Scholarship (2018.0371)'. We acknowledge funding within the framework of the NCCR PlanetS supported by the Swiss National Science Foundation grants no. 51NF40 182901 and 51NF40 205606. Dr. Harry Becker and Smithsonian Institution are thanked for providing Allende powder samples. Lorenz Gfeller from the Institute of Geography, University of Bern, is thanked for assistance with the ICP-MS analysis of the samples. We thank Dr. Gopalan Srinivasan for editorial handling and Dr. Elishevah van Kooten and an anonymous reviewer for their constructive comments that helped to improve the manuscript.

# Figures

**Figure 1**

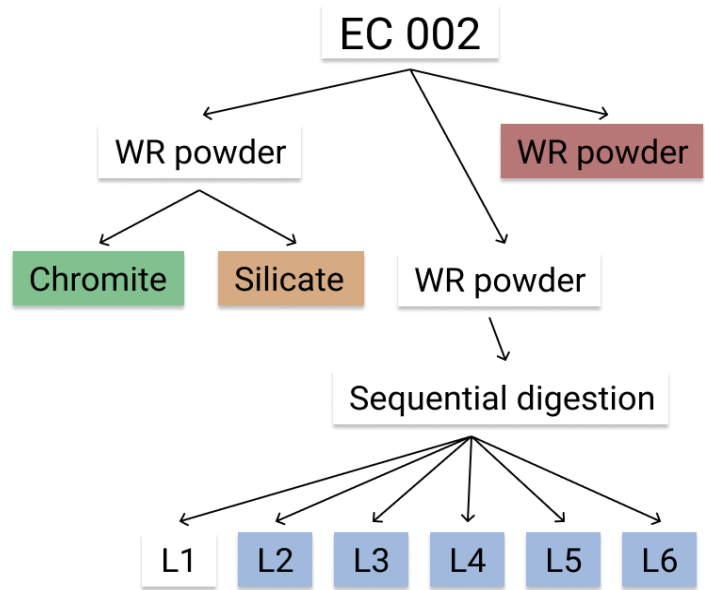

**Figure 1:** Separation scheme for EC 002 fractions. All fractions with coloured labels were analysed. WR: whole rock, L1-6: sequential digestion leachates.

**Figure 2**

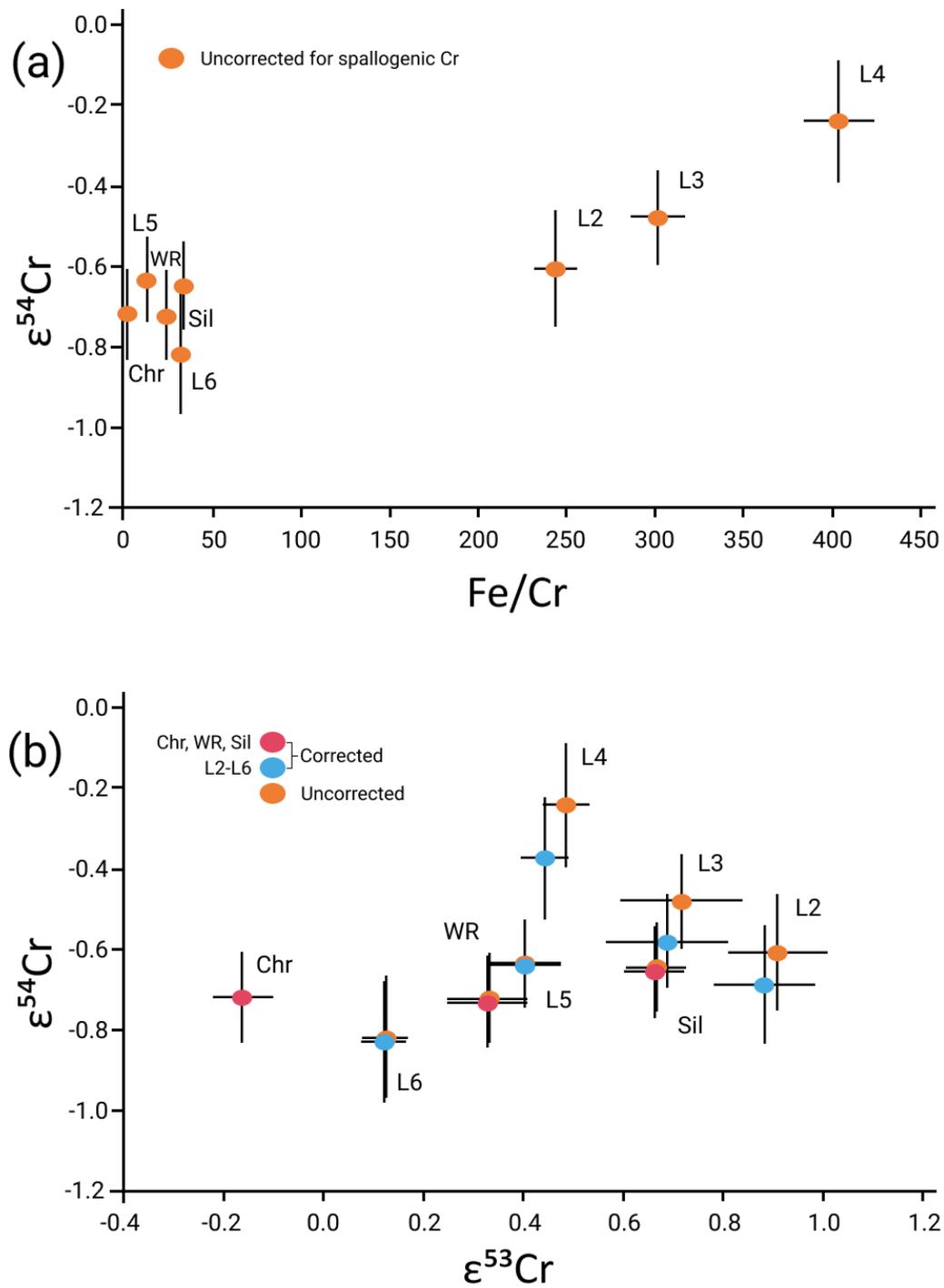

**Figure 2:** (a) $\varepsilon^{53}$Cr vs. Fe/Cr data for EC 002 fractions. Leachates 2, 3 and 4 show a correlation corroborating spallogenic Cr contribution. (b) Spallogenic Cr corrected and uncorrected $\varepsilon^{54}$Cr vs. $\varepsilon^{53}$Cr data for EC 002 fractions.

**Figure 3**

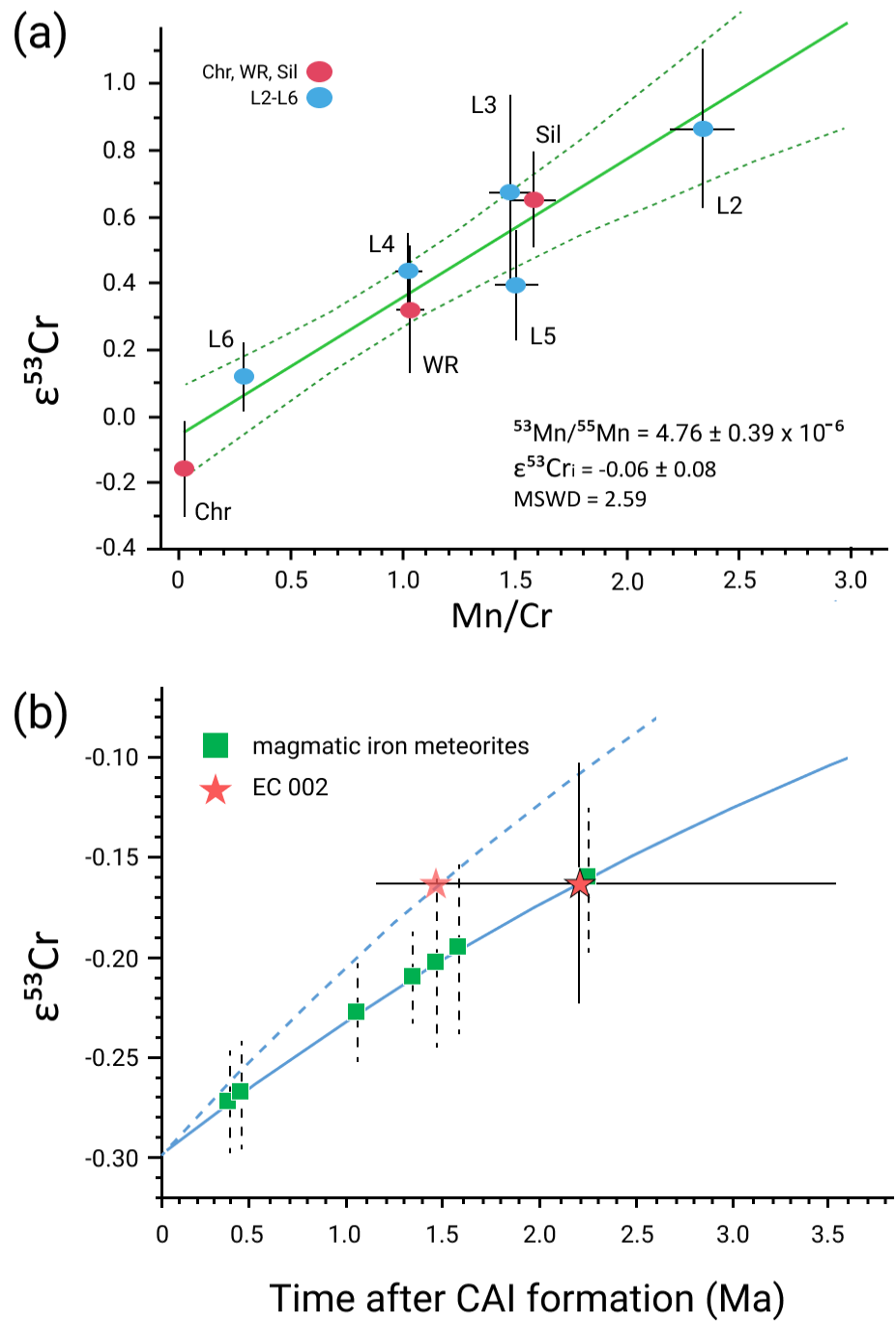

**Figure 3:** (a) $^{53}$Mn-$^{53}$Cr systematics of EC 002 fractions (Chr: chromite, WR: whole-rock, Sil: Silicates, L2-6: sequential digestion leachates). (b) $\varepsilon^{53}$Cr of EC 002 chromite fraction is plotted on a Cr isotope evolution curve corresponding to a chondritic reservoir (solid curve) and a reservoir with EC 002 whole rock-like $^{55}$Mn/$^{52}$Cr (dashed curve) through time (see text). $\varepsilon^{53}$Cr values of chromite/daubréelite fractions from magmatic iron meteorites are taken from Anand *et al.* (2021a). Error bars represent 2 s.e. uncertainties.

**Figure 4**

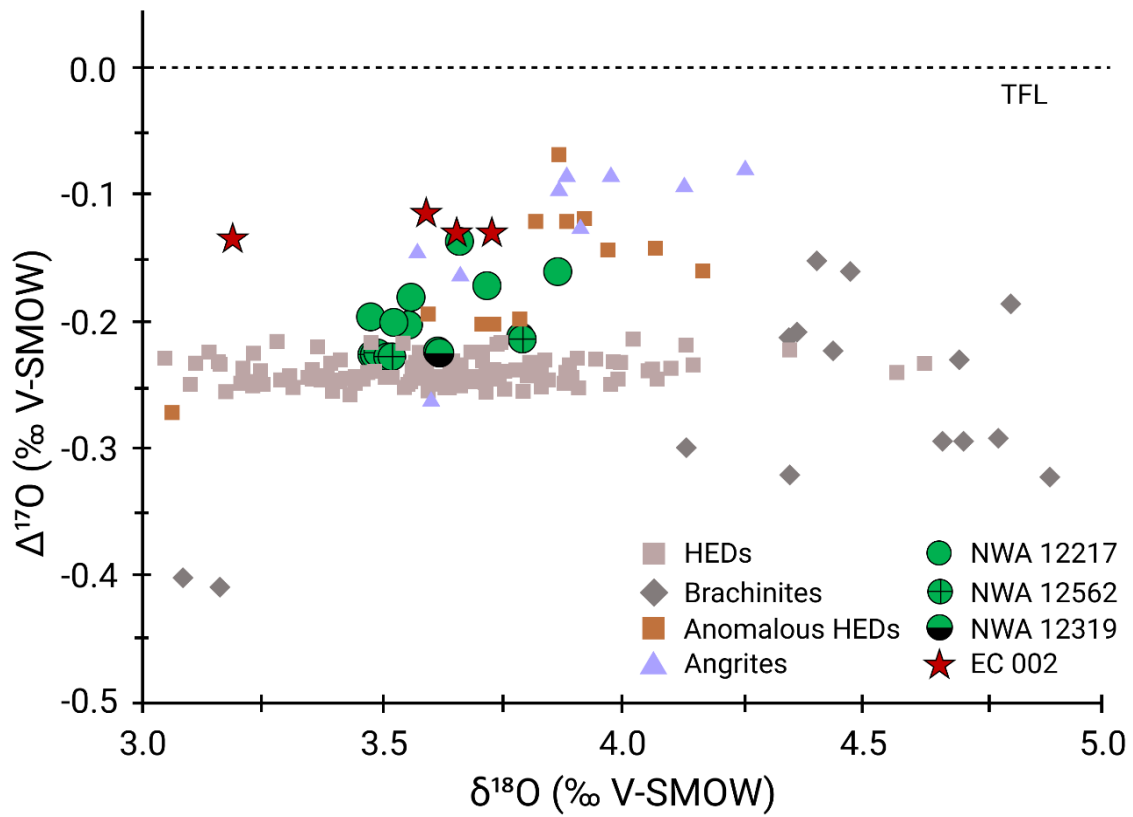

**Figure 4:** $\Delta^{17}O$ diagram showing EC 002, NWA 12217, 12319, and 12562 in relation to other achondrite groups and anomalous HEDs (Gattacceca *et al.*, 2021; Vaci *et al.,* 2021; Scott et al., 2009; Greenwood *et al.,* 2005, 2012; Zhang *et al.,* 2019). TFL: Terrestrial Fractional Line.

**Figure 5**

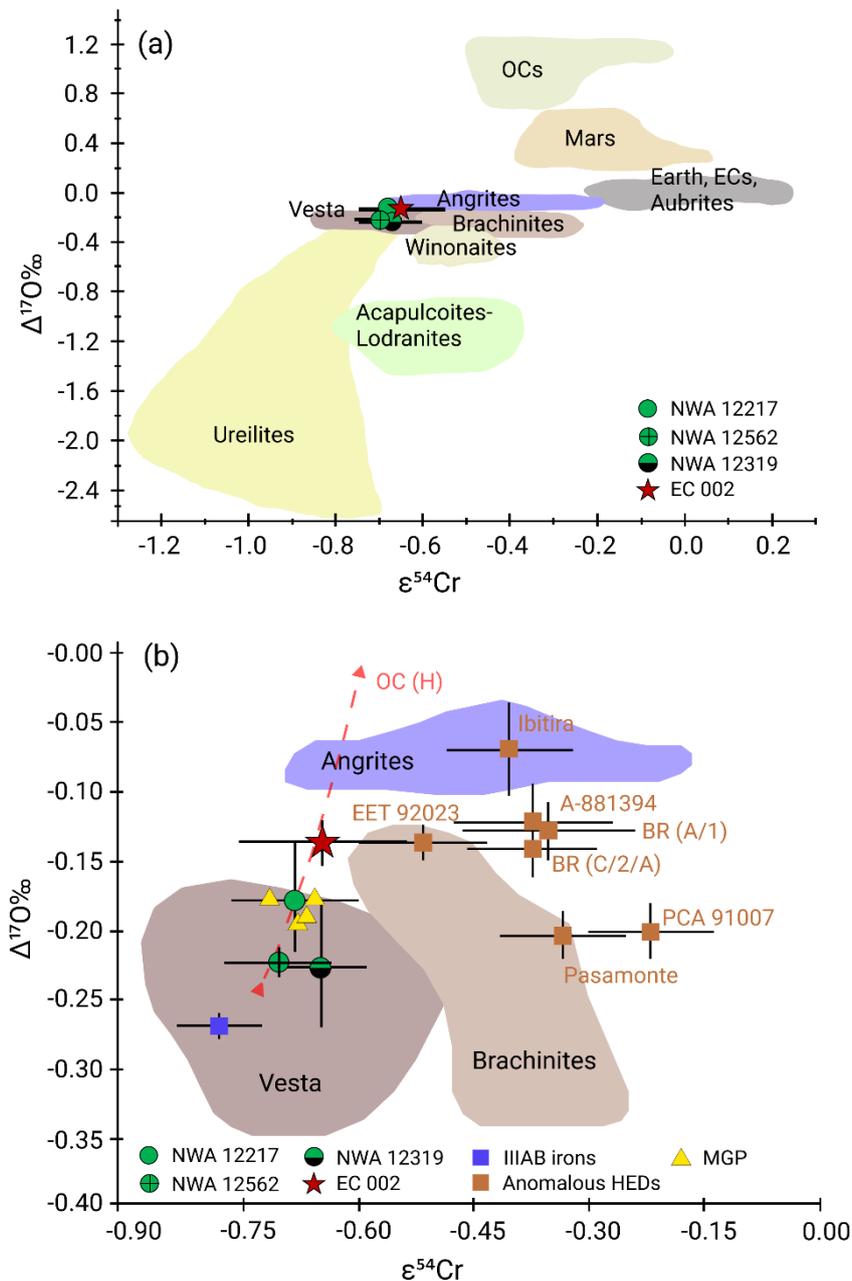

**Figure 5:** (a) $\Delta^{17}O$-$\varepsilon^{54}Cr$ diagram showing EC 002 with other achondrite and chondrite groups within 'non-carbonaceous (NC)' reservoir (OCs: ordinary chondrites, ECs: enstatite chondrites). Error bars for EC 002 represent 2 s.e. Literature data are from Sanborn *et al.* (2019) and references therein. (b) $\Delta^{17}O$-$\varepsilon^{54}Cr$ diagram showing EC 002 in comparison with brachinites, angrites, main group pallasites, IIIAB iron meteorites, normal and anomalous HEDs and NWA 12217, 12319, and 12562 ultramafic achondrites (Vaci *et al.,* 2021; Anand *et al.,* 2021a; Dey *et al.,* 2022). The dashed line represents mixing between normal HEDs with H ordinary chondrites.

**Figure 6**

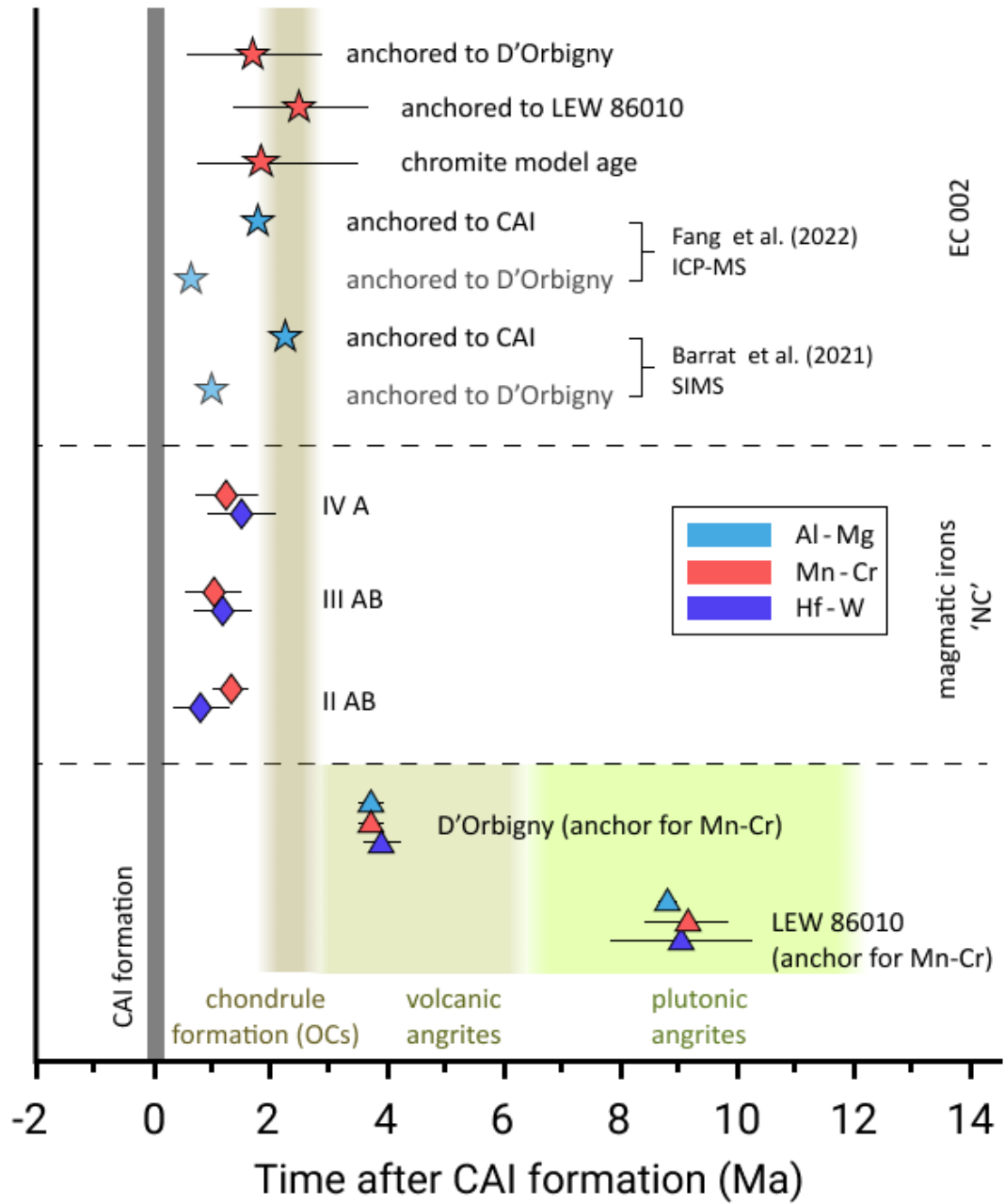

**Figure 6:** Compilation of chronological data for EC 002, angrites and magmatic iron meteorites. CAI formation age: Amelin *et al.*, 2010, Angrite ages: Zhu *et al.* (2019) and references therein, Iron meteorite ages: Anand *et al.* (2021a), Kruijer *et al.* (2017), EC 002 Al-Mg ages: Fang *et al.* (2022), Barrat *et al.* (2021), chondrule formation interval (OCs): Pape *et al.* (2019), Siron *et al.* (2021, 2022).

# Tables

**Table 1:** Sequential dissolution procedure and proportion of Mn and Cr leached in each sequential dissolution step for EC 002.

| Sequential digestion step | Reagent | Parameters (time/temperature) | Cr % | Mn % |
|---|---|---|---|---|
| L1 | 0.5 M $CH_3COOH$ | 0.5 h | 0.03 | 0.17 |
| L2 | 0.2 M $HNO_3$ | 0.5 h | 0.06 | 0.15 |
| L3 | 1 M HCl | 1 h | 0.18 | 0.28 |
| L4 | 6 M HCl | 48 h | 0.27 | 0.33 |
| L5 | conc. $HNO_3$ + HF (3:1) | 48 h, 140 °C | 62.31 | 88.85 |
| L6 | conc. $HNO_3$ + HF (3:1) | 70 h, 190 °C | 37.15 | 10.22 |

**Table 2:** $^{55}Mn/^{52}Cr$, Fe/Cr and Cr isotopic compositions of EC 002 fractions.

| Sample | $^{55}Mn/^{52}Cr$ | Fe/Cr | $\varepsilon^{53}Cr \pm 2se$ | $\varepsilon^{53}Cr^* \pm 2se$ | $\varepsilon^{54}Cr \pm 2se$ | $\varepsilon^{54}Cr^* \pm 2se$ | n |
|---|---|---|---|---|---|---|---|
| Whole rock | 1.04 | 24.36 | 0.329 ± 0.080 | 0.326 ± 0.080 | -0.718 ± 0.112 | -0.726 ± 0.112 | 4 |
| Silicates | 1.60 | 34.24 | 0.667 ± 0.060 | 0.664 ± 0.060 | -0.645 ± 0.112 | -0.656 ± 0.112 | 4 |
| Chromite | 0.03 | 1.27 | -0.163 ± 0.060 | -0.163 ± 0.060 | -0.716 ± 0.112 | -0.717 ± 0.112 | 3 |
| Leachate 1 | 7.32 | 239.09 | - | - | - | - | - |
| Leachate 2 | 2.36 | 243.29 | 0.908 ± 0.100 | 0.882 ± 0.100 | -0.607 ± 0.146 | -0.686 ± 0.146 | 4 |
| Leachate 3 | 1.49 | 301.93 | 0.717 ± 0.123 | 0.685 ± 0.123 | -0.478 ± 0.118 | -0.577 ± 0.118 | 5 |
| Leachate 4 | 1.03 | 403.95 | 0.485 ± 0.047 | 0.442 ± 0.047 | -0.240 ± 0.154 | -0.372 ± 0.154 | 3 |
| Leachate 5 | 1.52 | 12.32 | 0.404 ± 0.069 | 0.402 ± 0.069 | -0.634 ± 0.106 | -0.638 ± 0.106 | 3 |
| Leachate 6 | 0.29 | 32.44 | 0.123 ± 0.044 | 0.120 ± 0.044 | -0.817 ± 0.152 | -0.827 ± 0.152 | 4 |
| IAG OKUM[a] | 0.60 | 36.60 | 0.020 ± 0.065 | - | 0.083 ± 0.110 | - | 7 |
| Allende[b] | - | - | 0.072 ± 0.066 | - | 0.881 ± 0.120 | - | 4 |

Uncertainties associated with Cr isotopic compositions are reported as 2 s.e. of repeat analyses for each sample, or of NIST SRM 979 terrestrial Cr standard, whichever is larger. See Appendix for Cr isotopic composition of the individual runs. n = number of repeat measurements. The reproducibility of $^{55}Mn/^{52}Cr$ and Fe/Cr ratios is better than 5%. $\varepsilon^{53}Cr^*$ and $\varepsilon^{54}Cr^*$ represent the spallogenic Cr corrected data (and propagated uncertainties).

[a]IAG OKUM: Komatiite, Ontario (Certified Reference Materials, International Association of Geoanalysts), Peters and Pettke (2017).

[b]Allende rock standard (Smithsonian standard powder, USNM 3529, Split 18 position 1).

**Appendix:** Cr isotopic composition of EC 002 fractions and standard reference materials.

| Sample | | $\varepsilon^{53}Cr$ | s.e. int.[a] | $\varepsilon^{54}Cr$ | s.e. int.[a] | $\varepsilon^{53}Cr$ | | | $\varepsilon^{54}Cr$ | | |
|---|---|---|---|---|---|---|---|---|---|---|---|
| | | | | | | mean | 2 s.d. ext.[b] | 2 s.e. ext.[b] (n) | mean | 2 s.d. ext.[b] | 2 s.e. ext.[b] (n) |
| EC 002 Whole Rock | EC_WR_61_1 | 0.227 | 0.035 | -0.825 | 0.071 | 0.329 | 0.161 | 0.080 (4) | -0.718 | 0.316 | 0.112 (4) |
| | EC_WR_61_2 | 0.302 | 0.027 | -0.681 | 0.062 | | | | | | |
| | EC_WR_61_1b | 0.450 | 0.032 | -0.637 | 0.065 | | | | | | |
| | EC_WR_61_2b | 0.336 | 0.034 | -0.731 | 0.075 | | | | | | |
| Silicate | EC_Sil_61_1 | 0.607 | 0.035 | -0.782 | 0.071 | 0.667 | 0.169 | 0.060 (4) | -0.645 | 0.224 | 0.112 (4) |
| | EC_Sil_61_2 | 0.705 | 0.029 | -0.516 | 0.063 | | | | | | |
| | EC_Sil_61_1b | 0.726 | 0.031 | -0.553 | 0.069 | | | | | | |
| | EC_Sil_61_2b | 0.631 | 0.030 | -0.727 | 0.063 | | | | | | |
| Chromite | EC_Chr_61_1 | -0.155 | 0.035 | -0.796 | 0.075 | -0.163 | 0.169 | 0.060 (3) | -0.716 | 0.316 | 0.112 (3) |
| | EC_Chr_61_2 | -0.148 | 0.035 | -0.650 | 0.076 | | | | | | |
| | EC_Chr_61_1b | -0.185 | 0.036 | -0.703 | 0.078 | | | | | | |
| Leachate 2 | EC_L2_74_1 | 0.821 | 0.034 | -0.672 | 0.071 | 0.908 | 0.199 | 0.100 (4) | -0.607 | 0.293 | 0.146 (4) |
| | EC_L2_74_1b | 0.812 | 0.027 | -0.776 | 0.056 | | | | | | |
| | EC_L2_74_1c | 0.944 | 0.029 | -0.601 | 0.063 | | | | | | |
| | EC_L2_74_1d | 1.055 | 0.032 | -0.377 | 0.064 | | | | | | |
| Leachate 3 | EC_L3_74_1 | 0.579 | 0.029 | -0.602 | 0.066 | 0.717 | 0.274 | 0.123 (5) | -0.478 | 0.263 | 0.118 (5) |
| | EC_L3_74_1b | 0.828 | 0.031 | -0.226 | 0.067 | | | | | | |
| | EC_L3_74_1c | 0.530 | 0.029 | -0.557 | 0.061 | | | | | | |
| | EC_L3_74_1d | 0.869 | 0.030 | -0.504 | 0.060 | | | | | | |
| | EC_L3_74_1e | 0.781 | 0.030 | -0.503 | 0.055 | | | | | | |
| Leachate 4 | EC_L4_74_1 | 0.469 | 0.034 | -0.148 | 0.078 | 0.485 | 0.082 | 0.047 (3) | -0.240 | 0.267 | 0.154 (3) |
| | EC_L4_74_1b | 0.444 | 0.032 | -0.429 | 0.068 | | | | | | |
| | EC_L4_74_1c | 0.541 | 0.029 | -0.143 | 0.065 | | | | | | |
| Leachate 5 | EC_L5_74_1 | 0.328 | 0.030 | -0.644 | 0.064 | 0.404 | 0.120 | 0.069 (3) | -0.634 | 0.184 | 0.106 (3) |
| | EC_L5_74_2 | 0.408 | 0.030 | -0.742 | 0.065 | | | | | | |
| | EC_L5_74_2b | 0.475 | 0.029 | -0.517 | 0.062 | | | | | | |
| Leachate 6 | EC_L6_74_1 | 0.126 | 0.026 | -0.632 | 0.053 | 0.123 | 0.132 | 0.044 (4) | -0.817 | 0.303 | 0.152 (4) |
| | EC_L6_74_1b | 0.152 | 0.028 | -0.843 | 0.063 | | | | | | |
| | EC_L6_74_2 | 0.080 | 0.029 | -0.747 | 0.058 | | | | | | |
| | EC_L6_74_2b | 0.137 | 0.031 | -1.045 | 0.067 | | | | | | |
| IAG OKUM | OKUM_50_1 | 0.046 | 0.032 | 0.140 | 0.067 | 0.020 | 0.159 | 0.065 (7) | 0.083 | 0.270 | 0.110 (7) |
| | OKUM_50_2 | 0.034 | 0.038 | 0.204 | 0.081 | | | | | | |
| | OKUM_50_3 | 0.013 | 0.029 | 0.057 | 0.063 | | | | | | |
| | OKUM_61_1 | -0.054 | 0.029 | 0.114 | 0.064 | | | | | | |
| | OKUM_61_1b | -0.106 | 0.029 | -0.231 | 0.064 | | | | | | |
| | OKUM_61_2 | 0.168 | 0.031 | 0.173 | 0.063 | | | | | | |
| | OKUM_61_2b | 0.036 | 0.031 | 0.125 | 0.062 | | | | | | |
| Allende | All1 | 0.045 | 0.031 | 0.967 | 0.061 | 0.072 | 0.132 | 0.066 (4) | 0.881 | 0.241 | 0.120 (4) |
| | All1b | 0.092 | 0.030 | 0.865 | 0.063 | | | | | | |
| | All2 | -0.014 | 0.027 | 0.691 | 0.054 | | | | | | |
| | All2b | 0.167 | 0.031 | 1.001 | 0.070 | | | | | | |

The uncertainties associated with Cr isotopic compositions are reported as 2 s.e. of the replicate measurements of the samples or of the terrestrial Cr standard NIST SRM 979, whichever is larger.

[a]s.e. int.: internal error (reported as 1 s.e.) for a single filament run consisted of 24 blocks with 20 cycles each (integration time = 8.389 s).

[b]s.d. ext. and s.e. ext.: external errors of the replicate measurements reported as 2 s.d. and 2 s.e., respectively. n = number of replicate measurements.